\documentclass[showpacs,prl,twocolumn,floatfix]{revtex4}

\usepackage{graphicx,float,amsmath}

\begin{document}

\title{On giant piezoresistance effects in silicon nanowires and microwires}
\author{J.S. Milne}
\email{jason.milne@polytechnique.edu}
\affiliation{Physique de la mati\`ere condens\'ee, Ecole Polytechnique, CNRS, 91128
Palaiseau, France}
\author{S. Arscott}
\affiliation{Institut d'Electronique, de Micro\'electronique et de Nanotechnologie
(IEMN), CNRS UMR8520, Avenue Poincar\'e, Cit\'e Scientifique, 59652
Villeneuve d'Ascq, France}
\author{C. Renner}
\affiliation{Department of Condensed Matter Physics and NCCR Materials with Novel Electronic Properties, University of Geneva, 24 Quai Ernest-Ansermet, CH-1211 Geneva 4, Switzerland}
\author{A.C.H. Rowe}
\affiliation{Physique de la mati\`ere condens\'ee, Ecole Polytechnique, CNRS, 91128
Palaiseau, France}

\begin{abstract}
The giant piezoresistance (PZR) previously reported in silicon nanowires is experimentally investigated in a large number of surface depleted silicon nano- and micro-structures. The resistance is shown to vary strongly with time due to electron and hole trapping at the sample surfaces. Importantly, this time varying resistance manifests itself as an apparent giant PZR identical to that reported elsewhere. By modulating the applied stress in time, the true PZR of the structures is found to be comparable with that of bulk silicon.\end{abstract}

\pacs{73.50.Dn, 73.50.Gr, 73.63.Nm}
\maketitle

As the most well studied and commercially important semiconductor, reports of new physical phenomena in silicon receive much attention. A recent example is giant piezoresistance (PZR) \cite{he2006}, where the change in resistance of silicon nanowires due to an applied mechanical stress was reported to be orders of magnitude larger than that of bulk silicon \cite{smith1954}. This report is highly cited \cite{jie2008,neuzil2010,he2008,barwicz2010, reck2008, anderas2009} in part because it may represent another example of the effect of size on the physical properties of an otherwise well characterized material \cite{jie2008,delmo2009,hochbaum2008}. Additionally, giant PZR is currently seen as a potential breakthrough means of detecting motion in nano-electromechanical systems \cite{he2008} where conventional detection methods lose sensitivity \cite{mile2010,neuzil2010,ekinci2005}. Moreover, since mechanical stress is a key element for performance enhancement of microelectronic devices \cite{itrs2009}, the physical mechanism behind giant PZR could prove to be an important ingredient in the design of future nano-scale transistors. As yet there is no consensus concerning the origins of giant PZR, although two models have some support.\cite{cao2007,rowe2008} One \cite{cao2007} is based on a surface quantization effect predicted to occur in the first few silicon monolayers, while the other \cite{rowe2008} is based on a stress-induced movement of the surface Fermi level in partially depleted structures resulting from a change in surface charge. The atomic length scale of the former seems to be in disaccord with the typical wire diameters reported in the literature, which are at least several tens of nanometers, whereas the characteristic length scale of the latter is the surface depletion layer width (1 nm to 10 $\mu$m, depending on the doping density). It has been noted that the initially reported giant PZR occurred only in surface depleted nanowires \cite{he2006,rowe2008}, and subsequent claims of giant PZR involve depleted structures\cite{neuzil2010,barwicz2010,reck2008}. 

Here we show that in depleted structures resistance changes are dominated by electron and hole trapping at the surface. Quite unexpectedly this dielectric relaxation (which is independent of applied mechanical stress), results in apparent giant PZR signatures identical to those initially reported in silicon nanowires \cite{he2006}. The true PZR can only be measured with accuracy by modulating the mechanical stress in time as outlined below. In all cases it is comparable with that of bulk silicon \cite{smith1954}.

\begin{figure}[tbp]
\includegraphics[clip,width=8 cm] {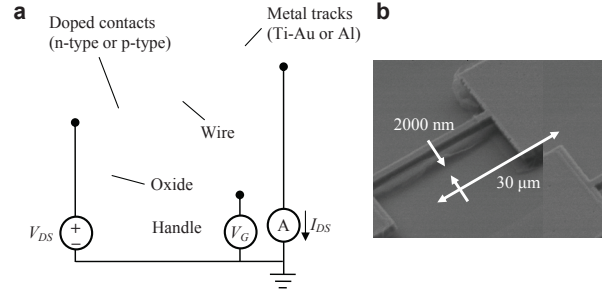}
\caption{(a) Typical layout of nanostructures, showing symbols used in the text and (b) SEM image of a released 2000 nm $\times$ 2000 nm $\times$ 30 $\mu$m microwire.}
\label{fig1}
\end{figure}

\begin{figure}[tbp]
\includegraphics[clip,width=8 cm] {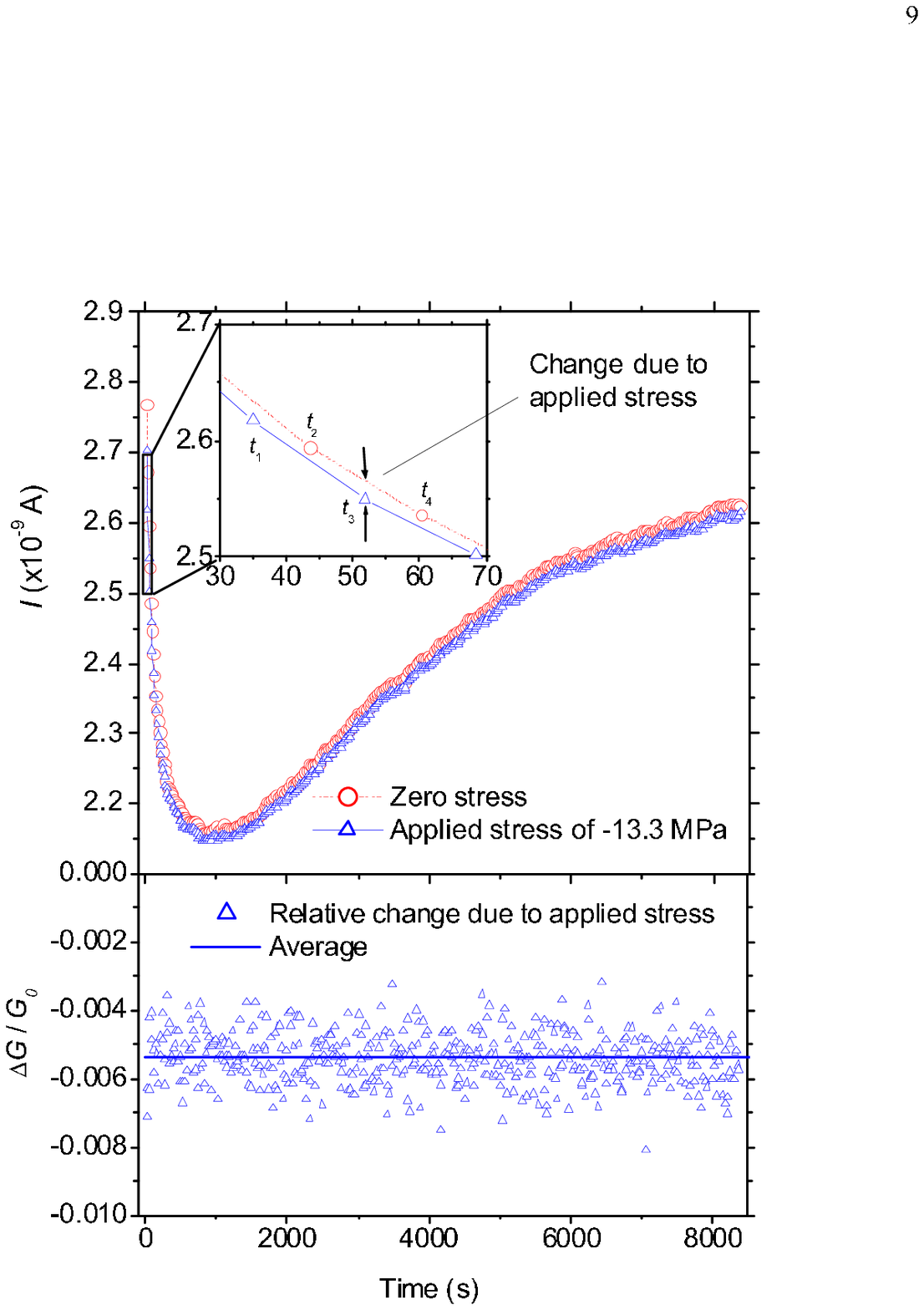}
\caption{(Top frame) Measurement of $I_{DS}(t)$ after applying a source-drain voltage of $V_{DS}$ = 0.5 V at $t = 0$ across a 200 nm $\times$ 2000 nm $\times$ 30 $\mu$m n-type nanoribbon and alternating the applied stress between 0 MPa and -13.3 MPa. $V_G$ was held at 0 V for the duration of the measurement. The inset indicates the sequence of measurements used in the stress modulation technique, where the lines are a guide to the eye. (Bottom frame) Relative conductance change due solely to the applied stress.}
\label{fig2}
\end{figure}

A variety of unreleased and released, n-type and p-type microwires, nanowires and nanoribbons were fabricated using a top-down approach from silicon-on-insulator wafers of different device layer thicknesses ($h$) and background doping levels (see Fig. \ref{fig1} and Table \ref{results}). The background doping density was chosen so that the surface depletion layer width \cite{sze2007} $W_D > h$, thereby ensuring that all structures are strongly depleted. All structures were etched using deep reactive ion etching and either thermally activated phosphorous (n-type) or boron (p-type) were used as dopants for the Ohmic contacts ($1\times10^{20}$ cm$^{-3}$). The metal contacts and tracks were composed of Ti/Au for n-type contacts and Al for p-type contacts. For released structures, the buried oxide (BOX) was removed using an HF (50 $\%$) etch followed by supercritical CO$_2$ drying to avoid possible wire stiction. All wires and ribbons are aligned with the $\langle 110 \rangle$ crystal direction along which the mechanical stress is applied using a three-point bending method. When resistance is measured in the same direction as the applied stress, $X$, the PZR is described by the longitudinal piezoresistive coefficient,
\begin{equation}
\pi_l=-\frac{1}{X}\frac{\Delta G}{G_0},  \label{picoeff}
\end{equation}
where $\Delta G$ is the change in conductance from a zero-stress value, $G_0$. In the $\langle 110 \rangle$ direction, $\pi_l = -32 \times 10^{11}$ Pa$^{-1}$ for bulk n-type silicon and $\pi_l = 72 \times 10^{11}$ Pa$^{-1}$ for bulk p-type silicon\cite{smith1954}. On wafer, large area strain gauges were also fabricated, thereby allowing the applied stress to be monitored in-situ. The three-point bending setup, which uses a piezoelectric pusher, permits rapid and repeatable switching between zero-stress and applied stress regimes; this experimental technique is different to the usual one where applied stress is ramped or stepped monotonically with time \cite{he2006,neuzil2010,barwicz2010,choi2008}. $G$ was measured by monitoring the current, $I_{DS}$, through the structures under an applied voltage, $V_{DS}$, while the silicon handle was held at a constant voltage back-gate voltage, $V_G$ (see Fig. \ref{fig1}).

\begin{figure}[tbp]
\includegraphics[clip,width=9 cm] {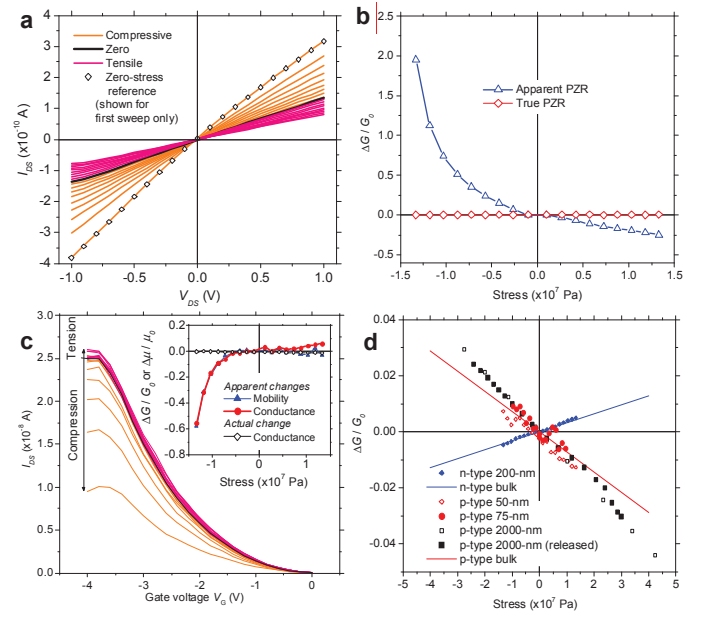}
\caption{(a) Successive $I_{DS}-V_{DS}$ measurements of a 50 nm $\times$ 50 nm $\times$ 1 $\mu$m p-type nanowire with $V_G$ = 0 V. (b) Apparent and true PZR extracted from the $V_{DS}$ = 0.5 V data points in (a). (c) Successive $I_{DS}-V_G$ measurements of a 50 nm $\times$ 50 nm $\times$ 1 $\mu$m p-type nanowire with $V_{DS}$ = 0.5 V, with an inset showing the apparent changes in mobility and conductance (calculations described in the methods section). (d) Relative conductance change due solely to the applied stress of various samples extracted using the stress modulation technique, together with typical values for n-type and p-type bulk silicon.}
\label{fig3}
\end{figure}

The need to separate the time-varying and stress-induced resistance changes can best be illustrated with the experimental data shown in Fig. \ref{fig2}. At $t = 0$, $V_{DS}$ is stepped from 0 V to 0.5 V across a 200 nm $\times$ 2000 nm $\times$ 30 $\mu$m nanoribbon with n-type contacts, while $X$ was alternated between 0 MPa and -13.3 MPa. $G$ reduces by 27 $\%$ during the initial 1000 s of measurement, and then increases for the remaining 7000 s, independent of $X$. Using the stress modulation technique, sequential measurements of $I_{DS}$ are made at zero stress (times $t_1$ and $t_3$) and with applied stress (times $t_2$ and $t_4$) (see inset). The true relative change in the conductance $\Delta G/G_0 = I_{DS}/I_{DS,0}$ due to the applied stress at $t_3$ is found by linear interpolation to be [$2I_{DS}(t_3) - I_{DS}(t_2) - I_{DS}(t_4)]/ [I_{DS}(t_2) + I_{DS}(t_4)$] which remains at a constant -0.54 $\%$ (see bottom frame, Fig. \ref{fig2}) over time. Using Eq. \ref{picoeff}, $\pi_l = -41 \times 10^{11}$ Pa$^{-1}$, in excellent agreement with the bulk value for $\langle 110 \rangle$ oriented n-type silicon \cite{smith1954}.  If the stress had been ramped linearly in time, the true PZR would have been masked by the non-stress-related drift of $G$ which is 10-100 times larger. Indeed, the implicit assumption when using a linearly-ramped stress technique is that the zero-voltage resistance remains constant for the entire measurement. As is clear from Fig. \ref{fig2}, this is not necessarily valid for depleted silicon structures that, as will be seen below, are sensitive to surface charging. Similar dynamic changes in the resistance of silicon nanowires have been previously reported \cite{jie2008,sacchetto2010,anderas2009,fujii1999,fujii1999b}.

Figure \ref{fig3}a presents the results of a measurement designed to highlight how temporal changes in zero-stress resistance manifest themselves as an apparent PZR. Each solid line represents a single $I_{DS}-V_{DS}$ sweep for a particular applied stress, with $V_{DS}$ swept from -1 V to 1 V in increments of 0.1 V. Two zero-stress measurements of $I_{DS}$ were taken before and after the applied-stress measurement at each applied voltage, and averaged to give an accurate zero-stress reference. The applied stress was incremented between each $I_{DS}-V_{DS}$ sweep, from -13.3 MPa up to 13.3 MPa, including an applied stress of 0 MPa. The applied-stress measurements presented in this figure closely resemble published experimental data produced as evidence for giant PZR (see Fig. 2b of Ref. \onlinecite{he2006}), where the slope of each $I_{DS}-V_{DS}$ curve changes in step with the applied stress. However, the zero-stress reference measurements (indicated as diamonds for the first I-V sweep in Fig. \ref{fig3}a) indicate that the true PZR is negligible, and that the change is due solely to a non stress-related, time-dependent change in wire resistance over the duration of the measurement. Fig. \ref{fig3}b, which shows the apparent PZR (open triangles, calculated using the value of current at an applied stress of 0 MPa; the black line in Fig. \ref{fig3}a), and the true PZR (open diamonds, measured with respect to the zero-voltage reference). For clarity, the real and apparent PZRs are only depicted for an applied voltage $V_{DS}$ = 0.5 V, but similar results are obtained at all voltages. The apparent PZR has an exponential dependence on $X$ like the curves obtained in Ref. \onlinecite{he2006} (see Fig. 2c of that article) although other forms (including those labelled C, Z and I in Ref. 1) are observed at different moments along relaxation curves of the type shown in Fig. \ref{fig2}. Similarly, changes that are even larger or of opposite sign are equally possible. $\pi_l$ obtained from the apparent PZR in Fig. \ref{fig3}b is $\pi_l = 450 \times 10^{11}$ Pa$^{-1}$ around $X = 0$ MPa, and without use of the stress modulation technique, this value is indistinguishable from the true PZR.

\begin{figure}[tbp]
\includegraphics[clip,width=8 cm] {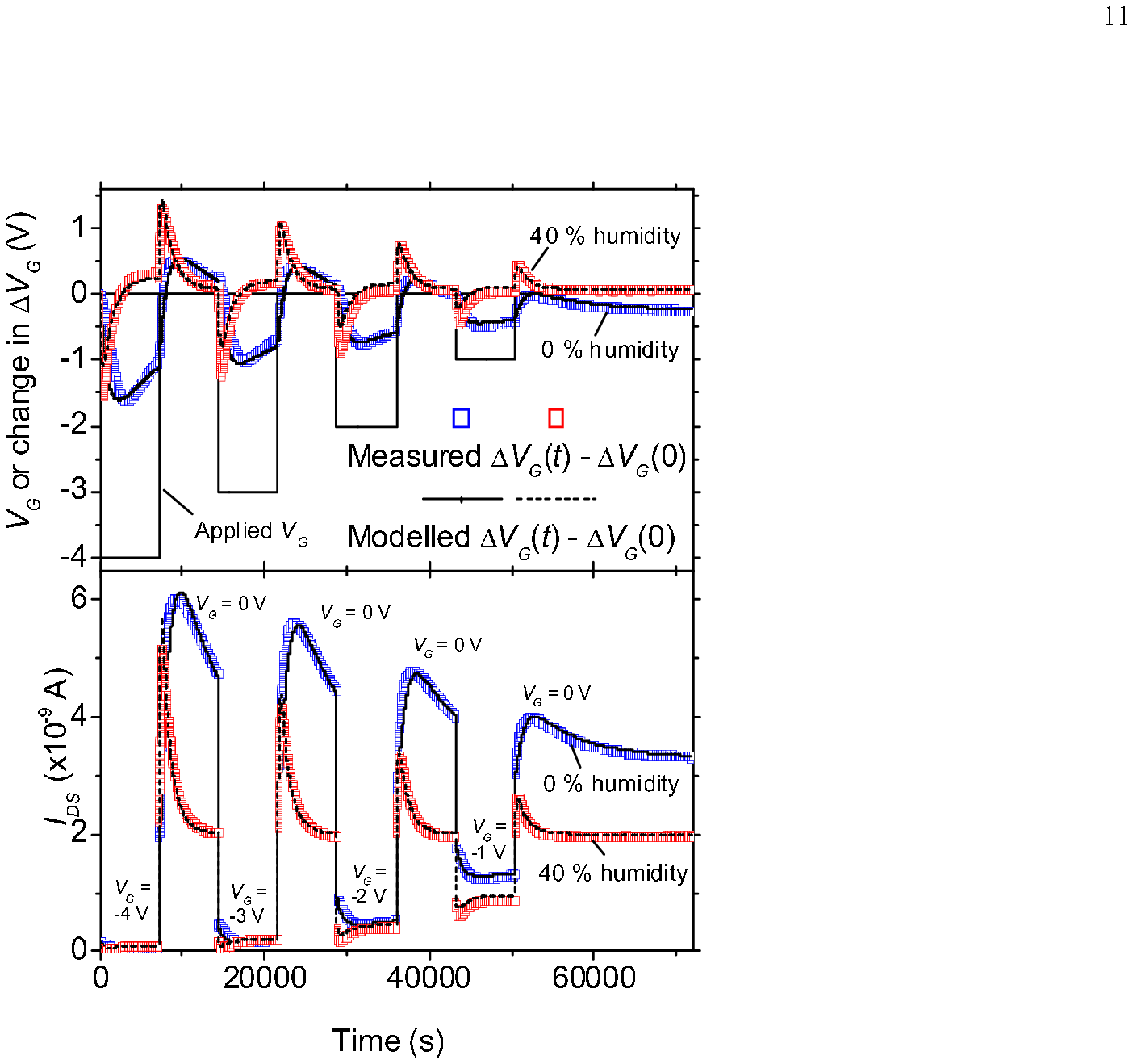}
\caption{Measurements and modelling of the change in $\Delta V_G$ due to oxide charge trapping and the values of $I_{DS}$ for a 200 nm $\times$ 2000 nm $\times$ 30 $\mu$m n-type nanoribbon, measured in 0 $\%$ and 40 $\%$ relative humidity. The source-drain voltage $V_{DS}$ was held constant at 0.2 V, while step changes were applied to the gate voltage $V_G$ as indicated in the top frame.}
\label{fig4}
\end{figure}

By measuring the change in $I_{DS}$ in response to a change in $V_G$, and by assuming a linear relationship between the slope of the $I_{DS}-V_G$ characteristic and the mobility, the giant PZR was attributed to a mobility variation.\cite{he2006}. This measurement is replicated using a 50 nm $\times$ 50 nm $\times$ 1 $\mu$m p-type nanowire. The results are presented in Fig. \ref{fig3}c, demonstrating that the apparent mobility extracted from the slope of an $I_{DS}-V_G$ measurement can also change over time independently of the applied stress. The sign and magnitude of this change is the same as the apparent stress-dependent conductance change. The true stress-dependent conductance change is consistent with bulk silicon PZR (see inset). Figure \ref{fig3}d shows the true values of $\Delta G/G_0$ as a function of stress obtained for four different, depleted silicon structures, together with the values expected from bulk silicon. Regardless of lateral wire size, or whether the device is released or not, the true PZR compares well with that of non-depleted, bulk silicon. Results from all measured samples can be found in Table \ref{results}.

To better understand the origin of the dynamic conductance changes, a similar approach to that used in a study of oxide traps in MOSFETs is employed, in which a measurement of the subthreshold current,
\begin{equation}
I_{DS} = I_0\exp[(V_G+\Delta V_G)/S],  \label{ids}
\end{equation}
is made.\cite{wang2002} Here $I_0$ is a constant, $S$ is the sub-threshold slope, and $\Delta V_G$ is the shift in the effective gate voltage due to trapped charge. $\Delta V_G$ at time $t$ can then be expressed as
\begin{equation}
\Delta V_G (t) - \Delta V_G (0) = S \ln\frac{I_{DS}(t)}{I_{DS}(0)}- V_G(t)+V_G(0), \label{deltavg}
\end{equation}
where $S$ can be measured directly from a rapid $I_{DS}$ versus $V_G$ measurement. Measured $V_G(t)-V_G(0)$ data can then be fitted using a semi-empirical model for positive and negative charge trapping in an oxide which is based on an existing model \cite{fischetti2009} for positive charge trapping in SiO$_2$ layers of metal-oxide-semiconductor (MOS) capacitors. The model  assumes that the sheet density $N$ of trapped charge has a saturation value $N_\infty$ that depends linearly on the electric field across the oxide $\xi_{ox}$, so that $N_\infty = \alpha \xi_{ox}$ where $\alpha$ is a constant with units of cm$^{-1}$V$^{-1}$. $N$ approaches $N_\infty$ with a time constant $\tau$, so that $\partial N/\partial t = -(N(t)-N_\infty)/\tau$, where $\tau$ is related to the capture cross-section of the trap and the current density through the oxide \cite{fischetti2009,young1979}. The contribution of the trapped charge to the gate voltage is given by \cite{fischetti2009} $\Delta V_G = qNd_{ox}/\epsilon_{ox}$, where $d_{ox}$ is the oxide thickness, $q$ is the charge on an electron, and $\epsilon_{ox}$ is the oxide permittivity. Figure \ref{fig4} presents measured values of $V_G(t)-V_G(0)$ for a 200 nm $\times$ 2000 nm $\times$ 30 $\mu$m n-type nanoribbon, where $V_{DS}$ = 0.2 V and a series of positive and negative steps were applied to $V_G$. Two sets of measurements were taken: one in a dry nitrogen atmosphere (0 $\%$ relative humidity) and one in a relative humidity of $\approx 40 \%$. Each set of data is modelled using one type of electron (hole) trap characterised by $N_e(0)$, $\alpha_e$ and $\tau_e$ ($N_h(0)$, $\alpha_h$ and $\tau_h$). Excellent agreement is reached between the model and the measured values of $V_G(t)-V_G(0)$ over each 20-hour measurement period. The calculated current is determined by using the modelled values of $V_G(t)$ in Eq. \ref{ids} and by adjusting the parameter $I_0$. This closely matches the measured current as shown in the bottom frame of Fig. \ref{fig4}. This treatment yields several important insights into the origin of the dynamic conductance changes: i) The observed changes in current are the result of charge trapping of electrons and holes in SiO$_2$ layers at the wires surfaces; ii)	The changes in current are much more rapid at higher relative humidity, consistent with the presence of water-related charge traps \cite{fischetti2009,nicollian1982}; iii)	The fitted values of the constant $\alpha_h$ for the hole traps ($1.2 \times 10^{12}$cm$^{-1}$MV$^{-1}$ at 0 $\%$ relative humidity and $1.5 \times 10^{12}$cm$^{-1}$MV$^{-1}$ at 40 $\%$ relative humidity) are almost identical to values obtained in MOS capacitor oxides \cite{fischetti2009}; iv)	Electron traps are associated with the rapid initial change in $I_{DS}$, while hole traps are associated with the slower change in the opposite direction, consistent with observations of charge trapping in MOS capacitors \cite{young1979}. This is strong evidence that the observed dynamic conductance changes are due to water-related charge traps in the oxide layer at the silicon surface. Consistent with this, the apparent giant PZR of Ref. \onlinecite{he2006} was also shown to depend strongly on the characteristics of this oxide layer. 

\begin{table}[tp]\footnotesize
\caption{True $\pi_l$ measured on all samples.}
\label{results}\centering 
\begin{tabular}{ccccccp{2.7cm}}

$h$ & $w$ & $l$ & $W_D$ & Doping & Released? & $\pi_l$ \\
(nm) & (nm) & ($\mu$m) & (nm) &  &  &   ($\times 10^{-11}$ /Pa) \\ \hline
2000 & 2000 & 30 & 8000 & p & no &  96.6, 90, 102, 96.8, 101, 107, 87.5  \\ \hline
2000 & 3000 & 30 & 8000 & p & no &  96.2, 102 \\ \hline
2000 & 2000 & 30 & 8000 & p & yes &  205, 115, 125 \\ \hline
2000 & 3000 & 30 & 8000 & p & yes &  116 \\ \hline
100 & 50 & 1 & 800 & p & no & 58.5 \\ \hline
75 & 50 & 1 & 800 & p & no & 74.6 \\ \hline
50 & 50 & 1 & 800 & p & no & 76.9 \\ \hline
200 & 2000 & 30 & 800 & n & no & -76, -77.4, -99, -48.9, -60.8, -46.3 \\ \hline
200 & 3000 & 30 & 800 & n & no & -69.7, -50.4 \\ \hline
\end{tabular}
\end{table}

In conclusion, charge trapping and de-trapping can mask the true PZR of depleted structures and mistakenly lead to claims of an apparent giant PZR. In more than 20 different surface depleted samples, the true PZR is found to be comparable with known values for bulk silicon (see Table \ref{results}).\cite{beaty2002} This is in stark contrast to previous reports of giant PZR in structures of similar dimensions and doping levels. While this does not rule out giant PZR in depleted silicon structures, future claims must conclusively demonstrate that any measured resistance change be solely due to the applied stress.

\bibliographystyle{apsrev}
\bibliography{achr}

\end{document}